\documentclass{emulateapj}
\usepackage{graphicx}
\input psfig.sty
\usepackage{multirow}
\usepackage{textcomp}
\usepackage{epsfig}
\usepackage{amsmath}
\usepackage{amssymb}
\usepackage{amsthm}
\usepackage{xy}
\def\sgr {SGR 0418+5729}
\def\ergs {${\rm erg\, s}^{-1}$}

\shorttitle{Is \sgr\ indeed a waning magnetar ?}
\shortauthors{Turolla et al.}

\begin{document}

\title{Is \sgr\ indeed a waning magnetar ?}

\author{R. Turolla\altaffilmark{1,2}, S. Zane\altaffilmark{2}, J.A. Pons\altaffilmark{3},
P. Esposito\altaffilmark{4}, N. Rea\altaffilmark{5}}
\altaffiltext{1}{Department of Physics, University of Padova, Via
Marzolo 8, I-35131 Padova, Italy} \altaffiltext{2}{Mullard Space
Science Laboratory, University College London, Holmbury St. Mary,
Dorking, Surrey, RH5 6NT, UK} \altaffiltext{3}{Department de
Fisica Aplicada, Universitat d'Alacant, Ap. Correus 99, 03080
Alacant, Spain} \altaffiltext{4}{INAF-Astronomical Observatory of
Cagliari, localit\`a Poggio dei Pini, strada 54, I-09012
Capoterra, Italy} \altaffiltext{5}{Institut de Ci\`encies de
l'Espai (CSIC-IEEC), Campus UAB, Facultat de Ci\`encies, Torre
C5-parell, E-08193 Barcelona, Spain}

\begin{abstract}

\sgr\ is a transient Soft Gamma-ray Repeater which underwent a
major outburst in June 2009, during which the emission of short
bursts was observed. Its properties appeared quite typical of
other sources of the same class until long-term X-ray monitoring
failed to detect any period derivative. The present upper limit on
$\dot P$ implies that the surface dipole field is $B_p\lesssim
7.5\times 10^{12}\ {\rm G}$ \cite[][]{rea10}, well below those
measured in other Soft Gamma-ray Repeaters (SGRs) and in the
Anomalous X-ray Pulsars (AXPs), a group of similar sources. Both
SGRs and AXPs are currently believed to be powered by
ultra-magnetized neutron stars (magnetars, $B_p\approx
10^{14}$--$10^{15}\ {\rm G}$). \sgr\ hardly seems to fit in such a
picture. We show that the magneto-rotational properties of \sgr\
can be reproduced if this is an aged magnetar, $\approx 1\ {\rm
Myr}$ old, which experienced substantial field decay. The large
initial toroidal component of the internal field required to match
the observed properties of \sgr\ ensures that crustal fractures,
and hence bursting activity, can still occur at present time. The
thermal spectrum observed during the outburst decay is compatible
with the predictions of a resonant compton scattering model (as in
other SGRs/AXPs) if the field is low and the magnetospheric twist
moderate.

\end{abstract}


\section{Introduction}\label{intro}

Soft Gamma Repeaters (SGRs) and Anomalous X-ray Pulsars (AXPs)
form a small, but rapidly expanding, class of isolated neutron
star (NS) sources characterized by the emission of short ($\approx
0.1$~s), energetic ($\approx 10^{40}$ \ergs) bursts of X-rays.
Their persistent luminosity ($L_X\approx 10^{32}$--$10^{36}$
\ergs\ in the $\sim 0.5$--10 keV range) is often variable, with
flux enhancements up to several hundreds during the outburst
phases of transient sources \cite[e.g.][]{reaesp11}. SGRs and AXPs
share similar timing properties, with periods in a narrow range
($P\sim 2$--12~s), and large period derivatives ($\dot{P} \approx
10^{-13}$--$10^{-10} \ {\rm s \ s}^{-1}$). In most of these
sources the X-ray luminosity exceeds the rate of rotational energy
losses ($\dot E\approx 10^{32}$--$10^{35}$ \ergs), and highly
variable radio activity has been detected in three objects. X-ray
spectra are often characterized by a thermal (BB; $kT\approx 0.5$
keV) plus a high-energy power-law (PL; photon index $\Gamma\approx
1.5$--3) component \cite[e.g.][]{rea08}, or by two thermal
components \cite[e.g.][]{gotthalp07,tiengo08}. Both the spectral
parameters and pulse profiles appear to vary with the source flux
\cite[see e.g.][for reviews on SGRs/AXPs properties]{woodsthomp06,
mereghetti08, reaesp11}\footnote{see also {\tt
http://www.physics.mcgill.ca/\textasciitilde
pulsar/magnetar/\\main.html} for an updated catalogue of
SGRs/AXPs}.

The large values of the magnetic field derived from the standard
dipole formula ($5\times 10^{13}\ {\rm G}\lesssim B_p \lesssim
10^{15}\ {\rm G}$), the lack of detected stellar companions and
their large X-ray output as compared to $\dot E$ led to the
suggestion that SGRs and AXPs are powered by a young (age $\approx
10^3$--$10^4$ yr), ultra-magnetized neutron star, or magnetar
\cite[][]{duncan92, thompson93}. Indeed, the magnetar scenario has
been largely successful in explaining many of the observed
properties of SGRs/AXPs, including those of the bursts
\citep{td95} and of the persistent emission \cite[the twisted
magnetosphere model,][and references therein]{thompson02, zrtn09,
alba10}.

Despite SGRs and AXPs are far from being a homogeneous class, in
particular the inferred surface dipolar field spans nearly two
orders of magnitude, their observational behaviour is now commonly
associated to that of (active) magnetars, to the point that often
the terms SGR/AXP and magnetar are used as synonyms. This,
actually, reflects the original definition of a magnetar as a NS
which is powered by its (large) magnetic field
\cite[][]{thompson93}. In this respect, a super-strong magnetic
field is not per se a sufficient condition for triggering
SGR/AXP-like activity, as testified by the existence of NS
sources, for instance most of the so-called High-B radio pulsars
\cite[HBPSRs; e.g.][]{kaspi10}, and possibly some of the thermally
emitting isolated NSs \cite[XDINSs; e.g.][]{turolla09}, with
surface magnetic fields comparable to those of SGRs/AXPs but
having substantially different properties and not showing any
bursting/outbursting activity over the $\sim 10$--20 yr time span
during which they were observed.

Over the last few years, the family of the magnetar candidates
grew with the addition of several new objects, most of them
transient. X-ray bursting activity or peculiar radio emission
similar to that of SGRs/AXPs was detected in allegedly
rotation-powered pulsars, as PSR J1846-0258 \cite[][]{gav08,
kum08} and PSR 1622-4950 \cite[][]{lev10}. This suggests that
magnetic energy substantially contributes to their emission at
certain stages. The magnetic field inferred from their rotation
properties ($B_p \sim 0.5$--$3\times 10^{14}\ {\rm G}$) is, in
fact, close to that of SGRs/AXPs, contributing to the widespread
belief that magnetar-like activity has to be associated to
super-strong magnetic fields, typically higher than the quantum
field $B_Q\simeq 4.4\times 10^{13} \ {\rm G}$.

In this respect, the recent discovery of a low-field SGR, \sgr\
\cite[][]{rea10}, came as a surprise. \sgr\ was observed for the
first time in June 2009 when it entered an outburst state during
which X-ray bursts were detected \cite[][]{vdhorst10}. The
enhanced flux level allowed for a measure of the source
periodicity, $P\sim 9.1$~s, but, despite the source was then
monitored in X-rays for $\sim 500$~d, no significant evidence for
a period derivative was found \cite[][and references
therein]{esp10, rea10}. The published upper limit is $\dot
P\lesssim 6\times 10^{-15}$ s$\,\rm s^{-1}$, leading to an
inferred dipole field $B_p\lesssim 7.5\times 10^{12}$~G.

The very low magnetic field of \sgr\ (when compared to other
SGRs/AXPs) raises a number of questions as to if, and how, the
observed phenomenology of this source can be accommodated within
the magnetar picture. A crucial point is whether a NS with a
surface field well below $10^{13}$~G, can harbor an internal
toroidal magnetic field strong enough to produce crustal
displacements, which are believed to be responsible for the
bursting/outbursting episodes in  SGRs/AXPs
\cite[][]{td95,thompson02, belob09}. In this paper we address this
issue and discuss both the spectral and timing properties of \sgr\
in the framework of an aging magnetar.

\section{SGR 0418+5729 as an old magnetar}\label{evol}

The suggestion that \sgr\ may be an aged magnetar was already put
forward by \citet[][see also \citealt{esp10}]{rea10} on the basis
of the low persistent luminosity (most likely well below that
observed during the outburst decay, $L_X\sim 6\times 10^{31}(D/2\,
{\rm kpc})$ \ergs), weak bursting activity \cite[only two faint
bursts were recorded,][]{vdhorst10}, and large spin-down age
($t_c\gtrsim 24\ {\rm Myr}$). \cite{rea10} estimated that an
internal toroidal field $B_{tor}\approx 5\times 10^{14}$~G is
required to power the persistent emission over a source lifetime
$\sim t_c$, and concluded that  $B_{tor}$ can be still large
enough to overcome the crustal yield.

The main appeal of the ``old magnetar'' scenario is that it can
offer an interpretation of the observed properties of \sgr\ within
an already established framework, validating the magnetar model
also for (surface) field strengths quite far away from those of
canonical SGRs/AXPs. However intriguing, the considerations
presented by \cite{rea10} necessary rely on quite crude estimates.
A more thorough investigation on the behaviour of evolved
magnetars, in which a substantial decay of the magnetic field
occurred, is definitely in order before claiming that \sgr\ is
indeed powered by the last hiccups of an once ultra-magnetized NS.

In the following we focus on three points which we deem central in
order to test the ``old magnetar'' hypothesis. First,  assuming
that \sgr\ was born with a magnetar-like surface field, $B_p$ must
have decayed by a factor $\approx 100$ to match the current upper
limit. Roughly the same reduction is expected in the internal
field. Although the latter can initially be $\approx 10$--100
times higher than $B_p$ (at least locally), one may wonder if at
late times internal magnetic stresses are still strong enough to
crack the crust. A second and related question is if realistic
models of field decay in magnetars can account for the observed
rotational properties (period and period derivative) of \sgr. This
also directly bears to the true age of the source which is most
probably much younger than the characteristic age, estimated
assuming a non-decaying field. A final point concerns the
persistent emission of \sgr. {\em XMM-Newton} observations show
evidence for a two component thermal spectrum (similar to those
observed in other transient magnetars), although the presence of a
non-thermal tail can not be excluded \cite[][see \S
\ref{persistent}]{esp10,esp11}. In the magnetar model, spectra are
expected to exhibit a power-law tail which originates from
resonant scattering in a twisted magnetosphere
\cite[][]{thompson02}. However, no calculations have been
performed yet in the range of surface fields implied by \sgr.

\subsection{Magneto-rotational Evolution}\label{magevol}

A major issue in establishing the magnetic evolution of NSs
(and of magnetars in particular) is that observations place very little,
if any, constraint on the structure and strength of the internal magnetic field.
While there are several indications that the large-scale, external field can be
reasonably assumed to be dipolar, with a moderate amount of twist in magnetars, the
different mechanisms proposed for the generation of the internal field
in the earlier phases of the NS life (differential rotation, dynamo, magneto-rotational
instability) most likely give rise to both  toroidal and poloidal components
\cite[e.g.][and references therein]{gep04,gep06}.
The presence of a toroidal field, roughly in equipartition with the poloidal one, is also
required by general stability arguments \cite[e.g.][and references
therein]{brait06}. A further complication comes from the present poor
knowledge of where the internal field resides. The field can either
permeate the entire star (``core'' fields), or be mostly confined in the crust
(``crustal'' fields), depending on where its supporting (super)currents
are located. The highly anisotropic
surface temperature distribution required in some XDINSs has
been taken as observational evidence in favor of a complex field geometry in the
external layers (crust, envelope, atmosphere) of NSs, either in the form of strong crustal
toroidal fields, multipolar poloidal components, or both \cite[][]{gep06,PMP06,zt06}.

The more general configuration for the internal field in a NS will
then be that produced by the superposition of current systems in
the core and the crust. As stressed by \cite{geppons07}, the
relative contribution of the core/crustal fields is likely
different in different types of NSs. In old radio pulsars, where
no field decay is observed, the long-lived core component may
dominate, while a sizeable, more volatile crustal field is
probably present in magnetars, for which substantial field decay
over a timescale $\approx 10^3$--$10^5$ yr is expected
\cite[e.g.][]{goldreis92}.

A particularly important result  \citep{Glam11} is the lesser role that ambipolar diffusion
plays in magnetar cores
(after crystallization, the absence of convective motions already quenched
ambipolar diffusion in the crust)
on their active lifetimes, contradicting an assumption often made in the modelling of the
flaring activity. Therefore, if the decay/evolution of the  magnetic field is indeed the cause
of magnetar activity, it is likely to take place outside the core and will be governed by
Hall/Ohmic diffusion in the stellar crust. The relative importance of these two mechanism is
strongly density- and temperature- dependent. Thus, any self-consistent study of the magnetic field
evolution must be coupled to a detailed modelling of the neutron star thermal
evolution, and conversely. Other mechanisms, e.g. flux expulsion from the superconducting core,
due to the interaction between neutron vortices and magnetic flux tubes,
are highly uncertain and very difficult to model.
For these reasons, recent investigations of the magnetic field evolution
in magnetars focused only on the crustal component of the field.

The first attempts in this direction used a split approach.
\cite{geppons07} studied the evolution of the field by solving the
complete induction equation in an isothermal crust, but assuming a prescribed time
dependence for the temperature. They found that crustal  magnetic
fields in NSs suffer significant decay during the first $\approx
10^6$~yr and that the Hall drift, although inherently conservative (i.e. alone it
can not dissipate magnetic energy), plays an important role
since it may reorganize the field from the larger to the smaller
(spatial) scales where Ohmic dissipation proceeds faster.

The cooling of magnetized NSs with field decay was investigated by
\cite{aguil08} by adopting a simple, analytical law for the time
variation of the field which incorporates the main features of the
Ohmic and Hall terms in the induction equation. The fully coupled
magneto-thermal evolution of a NS was finally addressed by
\cite{pons09}, including all realistic microphysics. However,
owing to numerical difficulties in treating the Hall term, their
models include only Ohmic diffusion. This can be a limitation
because, as they note, the Hall drift likely drastically affects
the very early evolution of ultra-magnetized NSs with surface
field $B_p\gtrsim 10^{15}$~G, and also that of ``normal'' NSs at
late times ($\gtrsim 10^6$~yr), when the temperature in the crust
has dropped. On the other hand, for initial values of $B_p\lesssim
10^{15}$~G, still well within the magnetar range, the effect of
the Hall drift is expected to introduce at most quantitative
changes (a somewhat faster dissipation) with respect to the purely
Ohmic picture.

\subsection{The Case of \sgr}

To explore if, and to which extent, the magneto-thermal evolution
of (initially) highly magnetic NSs can lead to objects with
properties compatible with those of \sgr, we performed some runs
using the code of \cite{pons09}. We refer to Section 2 in
\cite{pons09} and Section 4 of \cite{aguil08} for all details
about the code and the microphysical input. We evolved a
$1.4M_\odot$ NS assuming the minimal cooling scenario
\citep{page04}, with no exotic phases nor fast neutrino cooling
processes, but including enhanced neutrino emission from the
breaking and formation of neutron Cooper pairs in the NS core, as
recent observations of the Cassiopeia A supernova remnant seem to
require \citep{casa1,casa2}. The initial period was fixed at 10 ms
and the initial dipole field to $B_p=2.5\times 10^{14}$ G. Note
that the internal poloidal field is actually higher, with a
maximum value $B_{pol}(t=0)\simeq 2.5\times 10^{15}$ G in the
inner crust.

\begin{figure*}
\includegraphics[width=6.8in,angle=0]{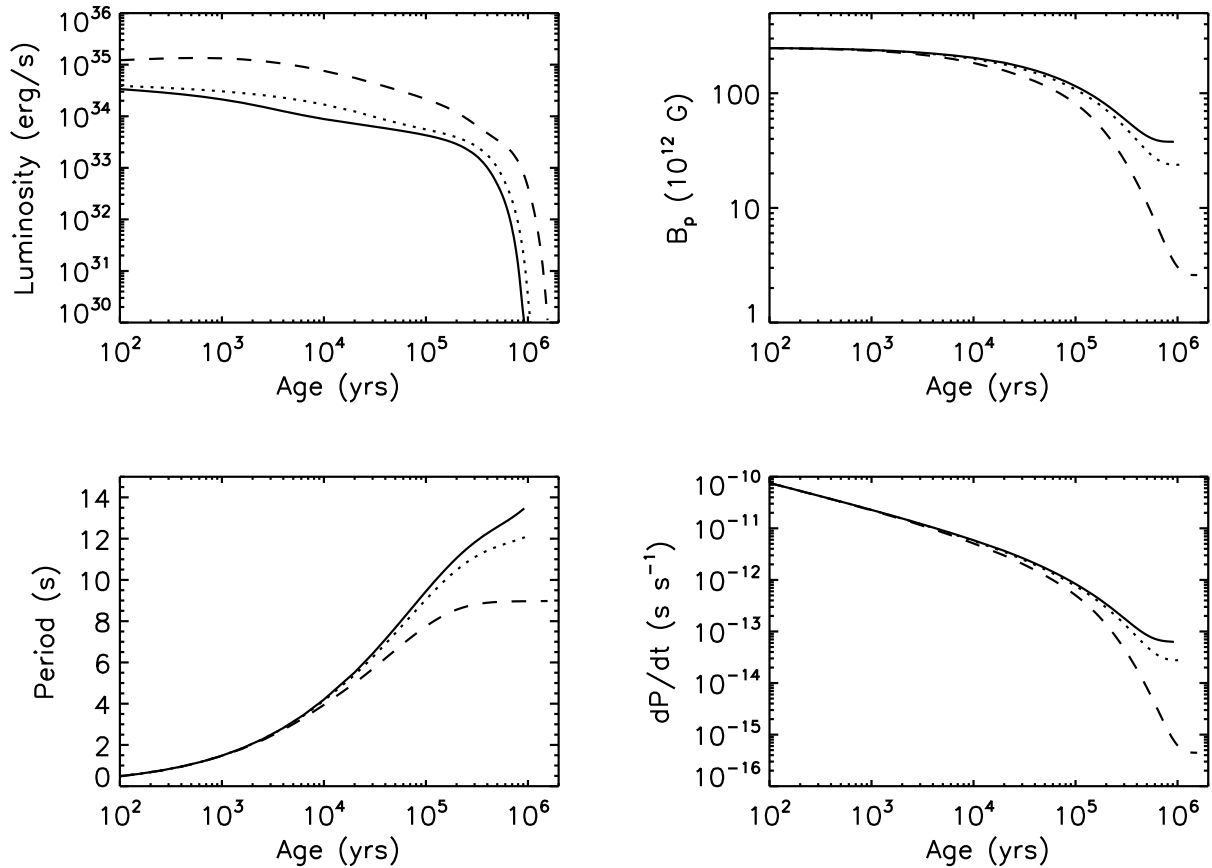}
\caption{From top left to bottom right, the evolution of the
luminosity, surface dipole field, period and period derivative
according to the model discussed in the text. The three cases
refer to  $B_{tor}(t=0)=0$ (solid lines), $B_{tor}(t=0)=4\times
10^{14}$~G (dotted lines) and $B_{tor}(t=0)=4\times 10^{16}$~G
(dashed lines). \label{mag-rot-evol}}
\end{figure*}

We considered three models with different values of the (maximum) internal toroidal field,
$B_{tor}(t=0)=0,\, 4\times 10^{14}$ and $4\times 10^{16}$ G, which turns out to be the crucial
parameter, as shown in Figure \ref{mag-rot-evol}.
The four panels illustrate the evolution of luminosity, dipole field $B_p$, period
$P$ and period derivative $\dot P$. Indeed the properties of \sgr\
are recovered in the case of $B_{tor}(t=0)=4\times 10^{16}$ G and
age $\sim 1.5 \times 10^6$ yr.  The main conclusions can be summarized as follows.

i) The low quiescent luminosity is easily explained considering
that the object is relatively old: even a NS born as a bright, hot
magnetar becomes cool and dim at this age.

ii) On the other hand, the observed period constrains the dipolar
field: the initial dipole field can not be much higher that the
one considered here in order to prevent the star from
spinning-down too fast and reach periods longer than that observed
at present. Obviously there are other large uncertainties, such as
the angle between rotation and magnetic axis, that may reduce the
period (we assumed an orthogonal rotator here).

iii) Although the components of the initial internal field
$B_{tor}(t=0)$ can be varied to some extent, a quite large value
is required. A large toroidal field, in fact, implies strong
currents, which, in turn, produce more heating and higher
temperatures. This drives a faster global field decay, which makes
it possible to match the observed upper limit on $\dot P$ and
$B_p$.

We stress that, while there are no stringent arguments against
such large internal fields in the NS crust, their real occurrence
in magnetars is an open issue. A possibility is that if the Hall
drift becomes very important and it results in much faster
dissipation, then one can obtain the same results starting with
lower initial toroidal fields. Finally, we note that the Hall term
is bound to become important again for objects like \sgr\ at late
stages ($\gtrsim 1$ Myr) as the star cools down and the
conductivity increases by several orders of magnitude. No
calculations are available in this regime but the possible
occurrence of a second ``Hall-active'' phase could lead to
enhanced bursting activity and rapid field decay. This may be an
indication that the estimate of the bursting rate in \cite{pern11}
is a lower limit.

\subsection{Occurrence of Bursts}\label{bursts}

Very recently \cite{pern11} used the magnetic evolution code of
\cite{geppons07} together with the cooling models by \cite{pons09}
to compute the magnetic stress acting on the NS crust at different
times. Their baseline model has $B_p(t=0)=8\times 10^{14}$~G and
$B_{tor}(t=0)=10^{15}$~G. They found that the occurrence of
crustal fractures (and hence of bursts) is not restricted to the
early NS life, during which the surface field is ultra-strong, but
can extend to late phases (age $\approx 10^5$--$10^6$ ~yr; see
their figure 2). Both the energetics and the recurrence time of
the events evolve as the star ages. For ``old'' magnetars about
$50\%$ crustal fractures release $\approx 10^{41}$~erg and the
waiting time between two successive events is $\approx 1$--10~yr.
They also made a longer run with a model with $B_p(t=0)=2\times
10^{14}$~G and $B_{tor}(t=0)=10^{15}$~G, for which the event rate
is about a factor 10 smaller.

The model we considered in \S \ref{magevol} as representative of
\sgr\ has $B_p(t=0)$ very close to this latter configuration,
while $B_{tor}(t=0)$ is larger. The present (maximum) value of the
internal toroidal field is $\sim 9\times 10^{14}\ {\rm G}$ and,
although we did not perform any detailed simulations, we argue
that the bursting rate of our model, at its present age, is
similar to the second model of \cite{pern11}, because the internal
configuration of the magnetic field is similar. It is important to
notice that, despite the much larger initial toroidal fields of
the model presented in this paper, this leads to faster decay and
therefore similar values are reached when the NS is a million
years old. Comparing both models at the estimated age of 1.5 Myr,
the internal toroidal field of the model presented in this paper
is only twice larger than that discussed in \cite{pern11}, and we
estimate the typical lapse time between events for an object like
\sgr ~ is $\sim 20$--50 years.

\subsection{Persistent Emission}\label{persistent}

In order to investigate the spectral properties of the persistent
emission from \sgr\ and its time evolution, we analyzed eight {\em
Swift} XRT\footnote{Each {\em Swift} dataset contains several
individual observations in order to obtain good enough
statistics.} and one {\em XMM-Newton} EPIC spectra \cite[see Table
S1 of][for more details]{rea10}. Preliminary results for some of
these datasets were already reported in \cite{esp10,esp11}.
Spectra for all the epochs were fitted simultaneously using XSPEC
v.12.6, with the value of the column density $N_H$ tied within the
different observations. Several one- and two-component models were
tried, including a single blackbody, a one-dimensional
\cite[RCS,][]{rea08} and a three-dimensional \cite[NTZ,][]{zrtn09}
resonant scattering model, a double blackbody and a blackbody plus
power-law. All single component models give rather poor fits.
While for the RCS and NTZ models the residuals and the $\chi^2$
values were not acceptable, a single BB decomposition properly
reproduces all the data but the {\em XMM-Newton} spectrum, which
is the only responsible for the relatively large $\chi^2$ of the
simultaneous fit. Since the highest-quality available spectrum
argues against the source having (at least at the early stage of
the outburst) a single thermal spectrum, we decided to add a
second component to the multi-instrument fit.

A BB+BB and a BB+PL model with all parameters free (except $N_H$,
see above) provide acceptable fits of comparable quality
($\chi^2_{red}=1.15$ for 601 dof and $\chi^2_{red}=1.12$ for 601
dof, respectively). However, we stress again that both these
spectral representations contain a large number (32 in total) of
free parameters, and those associated to the second component are
not required by the seven {\em Swift} observations but only by the
{\em XMM-Newton} one. Moreover, despite on a statistical ground
there is no reason to prefer the BB+BB over the BB+PL model, we
note that in the latter : i) the spectral index $\Gamma$ changes
dramatically and in a totally erratic way from one observation to
another, and ii) $\Gamma$ can be as large as $\sim 6$, arguing
against a power-law as a physically-motivated representation of
the second spectral component. On the other hand, the values of
the spectral parameters in the BB+BB model appear to be reasonable
and their time evolution is monotonic (see below).

Since in the BB+BB best fit model the temperature and
normalization of the colder BB component appear not to vary
sensibly in time (again, possibly because they are poorly
constrained by the {\em Swift} observations), we performed a fit
with these two parameters tied across the various datasets. This
resulted in a similarly good fit ($\chi^2_{red}=1.18$ for 617 dof)
and has the advantage to contain 16 degrees of freedom less. In
the case of the BB+PL model the goodness-of-the-fit worsens
considerably ($\chi^2_{red}=1.42$ for 617 dof) by requiring that
the parameters of the (single) BB are the same at the different
epochs.

For these reasons, in the following we take the BB+BB model (with
the colder BB constant in time) as the most likely representation
of the data, and discuss the ensuing implications in framework of
an evolved magnetar. It is worth mentioning that the failure of
the resonant scattering models to fit the data may be due to the
fact that both RCS and NTZ were originally developed for much
higher fields than that likely present in \sgr\ (the NTZ version
used here assumes $B_p=10^{14}$ G).  A more detailed spectral
analysis will be the subject of a forthcoming paper (Rea et al. in
preparation).

\begin{figure}
\includegraphics[width=3.4in,angle=0]{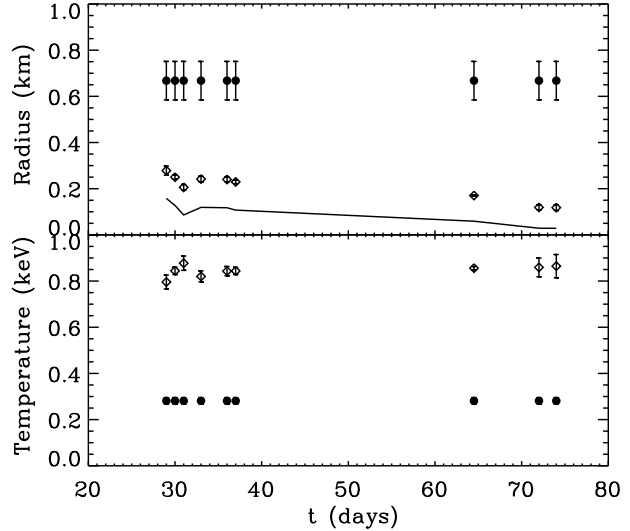}
\caption{The time evolution of the temperature and emitting radius
of the two BB components in the spectrum of \sgr; a source
distance of 2 kpc is assumed. Diamonds refer to the hot and filled
circles to the cold component. The solid line shows the ratio of
the emitting areas, $A_{hot}/A_{cold}$. Time is counted in days
from the outburst onset, on 2009-06-05 20:38:24.000 UTC (MJD
54987.862). \label{parevol}}
\end{figure}

The picture which emerges from the spectral analysis is that of
thermal emission from two regions on the star surface, a cold one,
with more or less constant size and temperature ($R_c\sim 0.75$~km
for a fiducial distance $D=2$~kpc and $T_c\sim 0.31$ keV), and a
hot one, which shrinks during the outburst decay. The evolution of
the temperature and size of the two components is shown in
figure~\ref{parevol}. The temperature of the hot region is more or
less constant at $kT_h\sim 0.93$ and its area changes from $\sim
0.2$ to $\sim 0.03$ times that of the cold region ($R_h\sim
0.15$--0.30~km, again for $D=2$~kpc). The overall behavior is
quite reminiscent of those seen in other transient magnetar
sources, notably the AXPs XTE J1810-197 and CXOU J164710.2-455216
\cite[e.g.][and references therein]{alba10}.

Within the magnetar model this is interpreted as due to the sudden
development of a twist in the external magnetic field which then
progressively decays. The twist likely affects only a limited
bundle of (closed) field lines and the charges flowing along the
current-carrying bundle heat the surface layers as they impact
upon the star. When the magnetosphere untwists, the size of the
heated region decreases \cite[][]{belob09}. This picture is
compatible with the results we obtained for \sgr\ with the BB+BB
model assuming that the heated region corresponds to the area
emitting the hotter BB. This is superimposed (or close) to a
cooler, larger cap which is responsible for the emission of the
softer BB. It is interesting to note that the analysis of the
pulse profiles during the first stages of the outburst supports
this view. The double-peaked pulse profile of \sgr\ suggests, in
fact, that the surface thermal map of the star comprises two warm
caps, only one of which was involved in the heating process
\cite[][]{esp10}. The predicted characteristic time for the
outburst evolution is $\approx 5 (\Phi/10^9\, {\rm V})^{-1}
(B/10^{14}\, {\rm G})\Delta\phi (A/10^{12}\, {\rm cm}^2)\ {\rm
yr}$, where $\Phi$ is the discharge voltage and $A$ is the area of
the surface region involved by the twist \cite[][]{belob09}.
Taking $B\sim 5\times 10^{12}\ {\rm G},\ \Delta\phi\sim 0.4$ rad
and $A\sim 10^{11}\ {\rm cm}^2$, we get for the characteristic
time $\approx 0.02(\Phi/10^9\, {\rm V})^{-1}{\rm yr}$. A low
discharge voltage, $\Phi\approx 10^8\ {\rm V}$, is then required
to obtain a decay time $\approx $ a few months. We warn that, as
already noted by \cite{esp10}, the luminosity produced by ohmic
dissipation appears to be too low to reproduce that observed at
the beginning of the outburst, $\sim 10^{34}(D/2\, {\rm kpc})^2$
\ergs. However, if the twist affected a region different from a
polar cap (e.g. a ring confined between two values $\theta_1$ and
$\theta_2$ of the magnetic colatitude) the value of the luminosity
can be higher\footnote{We thank A. Beloborodov for bringing this
point to our attention.}. Alternatively, other heating mechanisms
may be operating, e.g. the release of magnetic energy in the star
outer layers \cite[][]{let02}.

\begin{figure*}
\includegraphics[width=3.4in,angle=0]{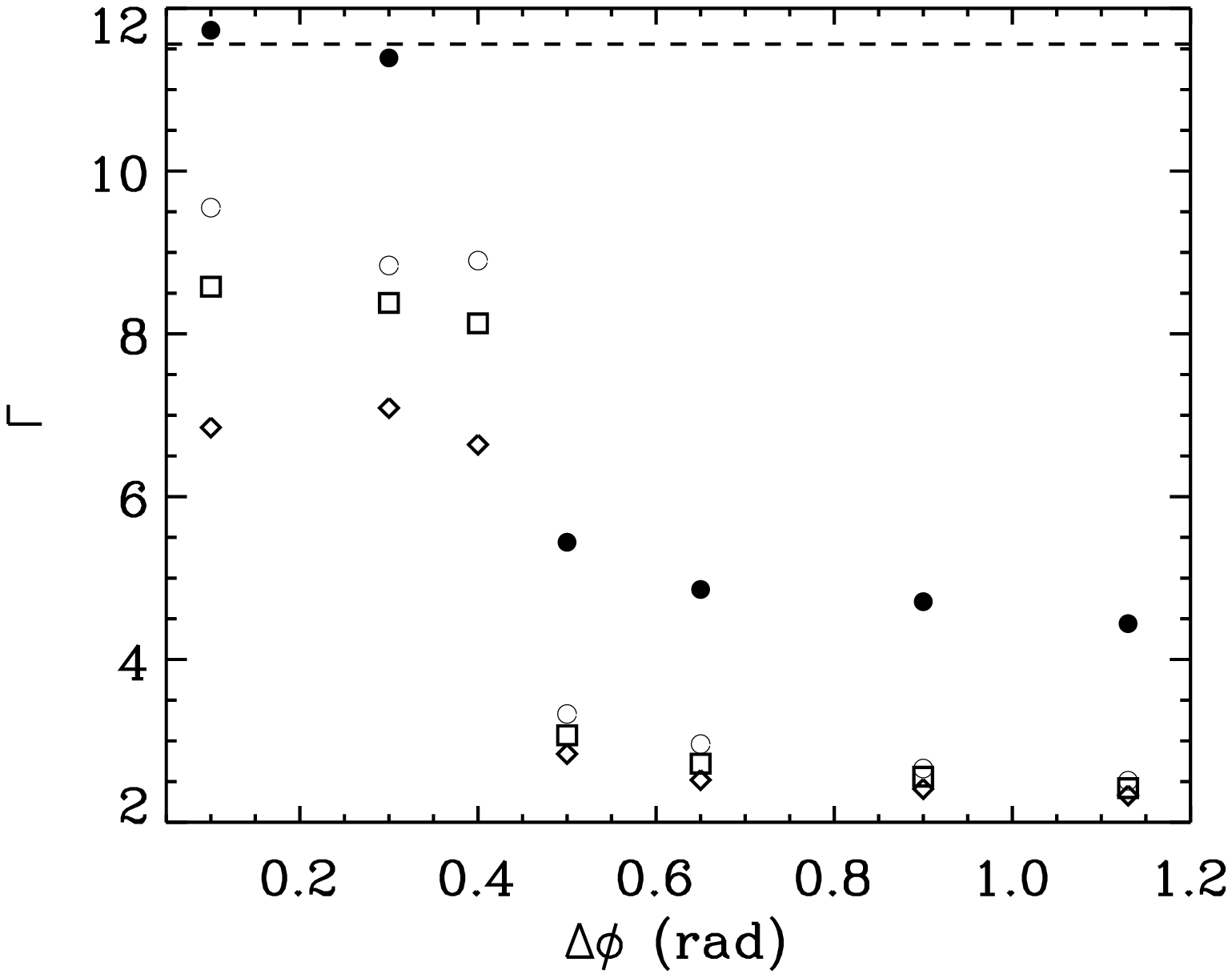}
\includegraphics[width=3.4in,angle=0]{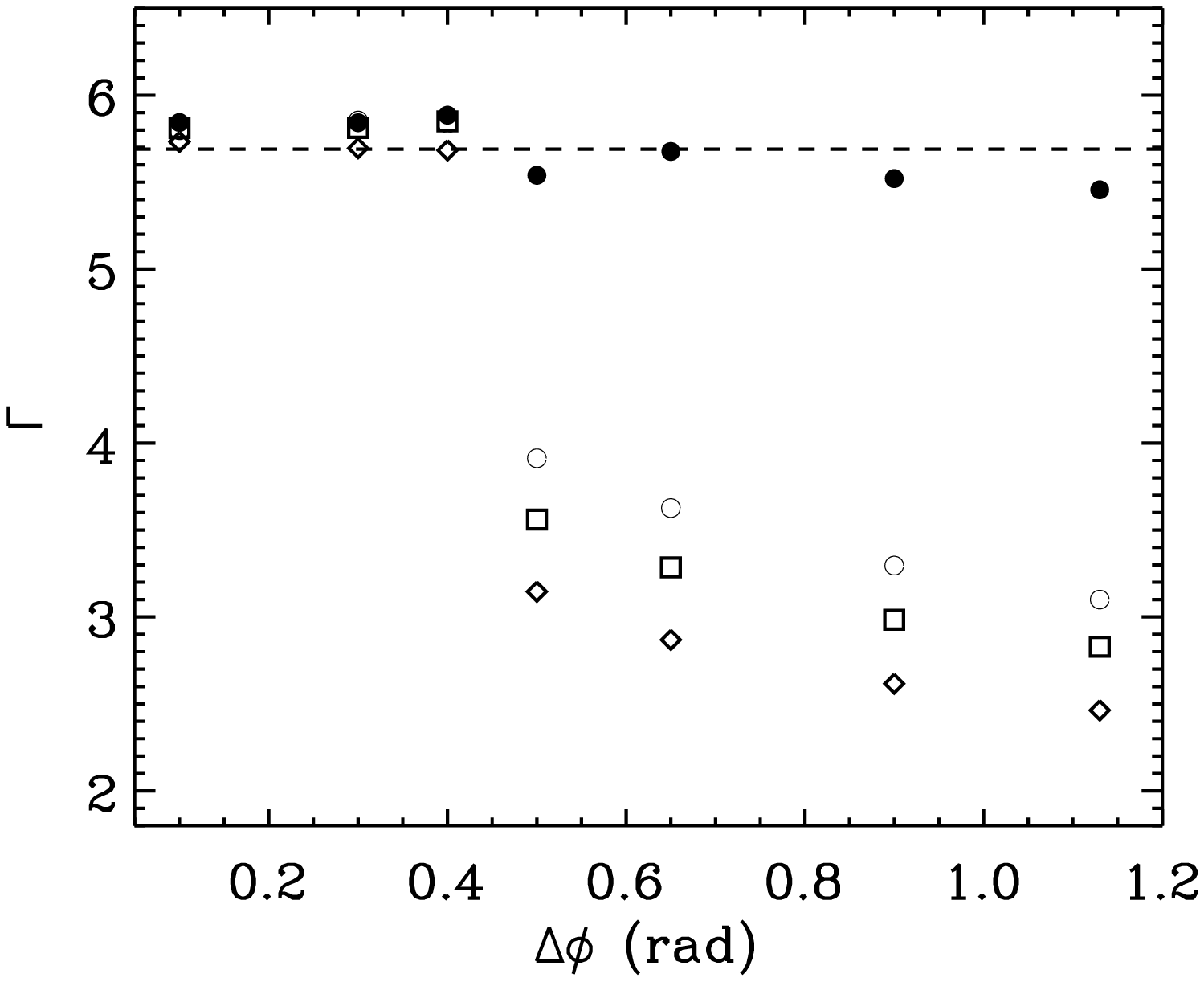}
\caption{{\em Left panel.} The photon index $\Gamma$ vs. the twist
angle $\Delta\phi$ for $B_p=10^{12}$ G (filled circles), $5\times
10^{12}$ G (open circles), $10^{13}$ G (squares) and $5\times
10^{13}$ G (diamonds); the seed photon temperature is $kT=0.3$
keV. The dashed horizontal line marks the value of the (average)
index of the blackbody spectrum at the same temperature and in the
same energy range (see text). {\em Right panel.} Same for $kT=0.9$
keV. \label{gamma}}
\end{figure*}

As discussed in Section \ref{bursts}, the internal field of \sgr\
can be still large enough to produce crustal displacements so that
magnetic helicity is transferred to the external field, twisting
up the magnetosphere. The appearance of a twist is usually
accompanied by the formation of a high-energy spectral tail, due
to resonant cyclotron up-scattering of thermal surface photons,
which is, however, not unambiguously detected in \sgr. In order to
investigate the properties of resonant cyclotron scattering
spectra in the low-field regime ($B_p\leq 5\times 10^{13}\ {\rm
G}$), we run a series of 3D Montecarlo simulations, using the
relativistic transport code of \citet[][]{nobili08a,nobili08b}, to
which we refer for all details. We considered four values of the
(polar) surface field ($B_p=10^{12},\, 5\times 10^{12},\,
10^{13},\, 5\times 10^{13} \ {\rm G}$), two values of the seed
photon temperature ($kT=0.3, \, 0.9\ {\rm keV}$) and several
values of the twist angle\footnote{Here the magnetosphere is
assumed to be globally twisted.} in the range $0.1\, {\rm
rad}<\Delta\phi<1.2\, {\rm rad}$. The electron temperature and
bulk velocity were fixed to $kT_{el}=10\ {\rm keV}$ and $v/c=0.5$
in all cases. A different choice of these parameters produce
similar results provided that the scattering particles are mildly
relativistic, as indeed required to reproduce the observed 1--10
keV spectra of SGRs/AXPs \cite[see e.g.][and references
therein]{nobili08a,nobili08b,zrtn09}. Results are summarized in
figure \ref{gamma} which shows the photon index of the non-thermal
tail (computed in the 6--8 keV range) as a function of the twist
angle for the different values of $B_p$. The (average) index (in
the same energy range) of the blackbody spectrum is marked by a
dashed horizontal line: when the photon index approaches the line
the spectrum becomes indistinguishable from a blackbody and no
tail is present. As it can be seen, while for $kT=0.3$ keV a
non-thermal tail below 10 keV appears for all the values of the
twist, unless $B_p=10^{12}$ G and $\Delta\phi \lesssim 0.3$ rad,
the up-scattering of seed photons associated to the hotter
component only produces a tail if $\Delta\phi \gtrsim 0.5$ rad and
$B_p>10^{12}$~G. We stress that here we are considering photon
energies below 10 keV, so the lack of a non-thermal spectral
component for $kT=0.9$ keV only reflects the fact that now
resonant comptonization tends to move photons at energies higher
than 10 keV. A tail, in fact, may be present above 10 keV also if
it does not show up below 10 keV.

\begin{figure}
\includegraphics[width=3.4in,angle=0]{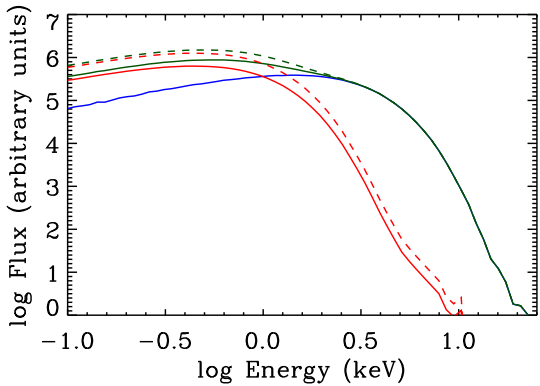}
\caption{The spectra obtained from the superposition of two
resonant scattering models with $kT=0.3,\, 0.9\ {\rm keV}$,
$\Delta\phi=0.4\, {\rm rad}$, $B_p=5\times 10^{12}$ G. The cold
component is shown in red, the hot in blue and the sum in green.
Solid (dashed) lines are for an emitting area ratio
$A_{cold}/A_{hot}=15$ (30). \label{spec}}
\end{figure}

Although the resonant scattering spectrum produced by the
reprocessing of soft photons coming from two NS surface regions at
different temperature is not exactly given by the superposition of
the two individual models \cite[see the discussion in][]{alba10},
we adopt this approach to get some insight into the spectral
properties of \sgr. If the observed spectrum of the source is best
modelled in terms of the superposition of two blackbodies with
$kT\sim 0.3,\, 0.9$ keV the twist angle must be $\lesssim 0.5$~rad
for $B_p>10^{12}\ {\rm G}$ not to produce a power-law tail in the
hot component (see figure \ref{gamma}). This, however, is only a
necessary condition because a PL tail may still appear in the cold
component. The emergence of such a power-law is related to the
relative magnitude of the hot and cold components. The total
spectrum resulting from the superposition of the models with
$kT\sim 0.3$ and 0.9 keV is shown in figure \ref{spec} for
$B_p=5\times 10^{12}$ G and two values of the ratio of the
emitting areas, $A_{cold}/A_{hot}=15,\, 30$, typical of those
measured during the evolution of \sgr. We remark that the largest
area ratio ($=30$) corresponds to the most unfavorable case: if
the tail does not appear now it is not present for smaller values
of the area ratio, when the cold component contributes less. As it
is seen, the total spectrum is very close to the superposition of
two blackbodies, with no high-energy tail. The same result holds
for different values of the magnetic field, provided that
$\Delta\phi\lesssim 0.5$ rad, and for even larger values of the
twist if the field is as low as $10^{12}$ G. We conclude that the
strong evidence that \sgr\ exhibits a X-ray spectrum dominated by
surface thermal emission below 10 keV is not in contradiction with
the predictions of RCS models.

\section{Discussion}

\sgr\ is the first ``low-field'' soft gamma repeater/anomalous
X-ray pulsar ever discovered. Even if other sources of the same
class with a relatively weak $B$-field were already known
\cite[e.g. 1E 2259+586,][]{gk02}, the case of \sgr\ is extreme and
stimulated considerable interest. With a surface dipole field
$\lesssim 7.5\times 10^{12}\ {\rm G}$, \sgr\ seems to challenge an
interpretation in terms of the ``conventional'' magnetar model,
and not just because its magnetic field is below the quantum
critical field, a quite irrelevant fact per se.

The present upper limit on the surface field in \sgr\ is based on
the spin-down measure, which traces the dipole component. It is
actually possible that the external magnetic field in NSs, and in
SGRs/AXPs in particular, is more complex than a simple dipole.
Higher order multipoles can substantially contribute to the field
near the surface and, because they fall off more rapidly with
radius, induce negligible spin-down. \sgr\ may, then, possess a
much higher surface $B$-field than indicated by its spin-down
rate.

A high surface field in the form of multipolar components hints
toward the presence a large internal field which can produce
crustal motions and bursting activity, making \sgr\ not dissimilar
(a part from the magnetic field topology) from other magnetar
sources. It does not, however, explain the rotational properties
of the source. With a dipole field below $10^{13}\ {\rm G}$, in
fact, it would be impossible to slow down the star to the observed
9.1 s period in less than $\sim 24$ Myr, a time much longer than
the estimated age of other SGRs/AXPs, unless \sgr\ had an
exceptionally long period at birth. A possible solution was
suggested by \cite{alpar11} who have shown that a if \sgr\ was
born with a period $> 70$ ms and a low dipolar field ($B\approx
10^{12}\ {\rm G}$), the torque exerted by a fallback disk can spin
down the star to the present period in $\gtrsim 10^5$ yr, if $\dot
P$ has to be below the observed upper limit\footnote{A somehow
similar scenario in which SGRs/AXPs are NSs with a low dipole
field and super-strong multipolar components powered by accretion
from a fallback disk was recently proposed by \cite{true10}.}. The
survival of multipolar field components $\approx 100$ times
stronger than the dipole over such a time span may, however, be an
issue in the light of the known evolution of the dipolar field in
magnetars (see \S \ref{magevol}).

In this paper we explored a different possibility, i.e. that \sgr\
is an old neutron star born with a super-strong magnetic field
which experienced field decay over a time $\approx 10^6$ yr. Our
results show that magneto-dipolar braking can effectively spin
down the star to a period $\sim 10$ s with $\dot P\lesssim
10^{-15}\ {\rm s\,s}^{-1}$ provided that the initial internal
toroidal field is large enough, $B_{tor}(t=0)\gtrsim 10^{16}\ {\rm
G}$. At the same time the initial external dipole field has to be
$\lesssim 2$--$3\times 10^{14}\ {\rm G}$ in order to prevent the
NS from spinning down too fast. In particular, the model with
$B_{tor}(t=0)=4\times 10^{16}\ {\rm G}$ and $B_{p}(t=0)=2.5\times
10^{14}\ {\rm G}$ gives $B_p\sim 2\times 10^{12}\ {\rm G}$, $P\sim
9$ s and $\dot P\sim 4\times 10^{-16}\  {\rm s\,s}^{-1}$ at an age
of $\sim 1.5 $ Myr, in very good agreement with the current
observational picture of \sgr. The predicted quiescent luminosity
at the same age is $L_X\sim 10^{31}$ \ergs, again in agreement
with the  current luminosity of the source, the faintest measured
so far and possibly close to the quiescent value \cite[$L_X\sim
6\times 10^{31}$ \ergs\ for a distance of 2 kpc;][]{rea10}. We
note in this respect that the luminosity drops very quickly for
ages $\gtrsim 10^6$ yr (top left panel of figure
\ref{mag-rot-evol}), so the present value could well be below
$\sim 10^{31}$ \ergs\ if the age is only slightly above 1.5 Myr. A
somehow longer age would have negligible impact on the predicted
$P$, $\dot P$ and $B_p$ which already reached a nearly constant
value (see again figure \ref{mag-rot-evol}). We also note that the
measured flux in the $\sim 0.5$--10 keV band may not be
representative of the bolometric luminosity for an old object like
\sgr. If the NS surface temperature is $\lesssim 10^6$ K, in fact,
the quiescent thermal emission would be too soft (and absorbed) to
be clearly detectable. It is then possible that the observed X-ray
flux only constrains the ``outburst'' luminosity and the genuine
quiescent emission may go undetected even if it is $\approx
10^{31}$ \ergs.

Given the large internal toroidal field at birth, our fiducial
model for \sgr\ retains the capability to induce crustal
fractures, and hence to produce bursts, even at quite late times.
A comparison of our model with one of the cases investigated in
detail by \cite{pern11} at a comparable age ($\sim 1.5$ Myr),
indicates that the recurrence time between bursting/outbursting
events for an object like \sgr\ is a few tens of years. Crustal
displacements are accompanied by a twisting up of the
magnetosphere and the formation of a high-energy spectral tail,
which is observed in most SGRs/AXPs. We have shown, however, that
a source with a spectral distribution consistent with a double
blackbody, as follows from  the analysis of a series of X-ray
observations taken during the outburst decay of \sgr, can be
modelled as well by a resonant cyclotron scattering model if the
twist is moderate (twist angle $\lesssim 0.5$ rad) and the surface
field is low, $\lesssim 5\times 10^{12}\ {\rm G}$, as predicted by
our evolutionary calculations.

It has been already noticed that the SGRs/AXPs which exhibit the
larger flux variations seem to be those with the lower dipole
fields \cite[][]{esp09}. \sgr\ provides a further case for such a
correlation. In the magnetar scenario the occurrence and
energetics of outbursts are dictated by the internal field and the
peak luminosity $L_{max}$ depends on the (maximum) energy stored
locally prior to the event. $L_{max}$ is not much sensitive to the
external dipole field. If the energy released in an event is
roughly similar in all sources (as in the case the mechanism is
gated e.g. by the crustal yield), the ratio between the peak and
persistent fluxes has to be much higher in the less active, old
objects than in active, young ones. This is because $L_{max}$ is
similar for all of them, but the quiescent luminosity is much
lower for old sources. In the latter, field decay had the time to
reduce the dipole field to rather low values, so old, low-field
objects (but with a sufficiently high internal field) appear as
transient sources. It is intriguing that \sgr\ is the source with
the lowest field known and at the same time a most extreme
transient, with a peak-to-persistent flux ratio $\sim 1000$.

\acknowledgements This research was
partially funded through grants AAE I/088/06/0 (RT), AYA
2010-21097-C03-02, GVPROMETE02009-103 (JP) and AYA2009-07391,
SGR2009-811, TW2010005 (NR). NR also acknowledges support from a
Ramon y Cajal Fellowship. PE acknowledges financial support from
the Autonomous Region of Sardinia through a research grant under
the program PO Sardegna FSE 2007-2013, L.R. 7/2007 ``Promoting
scientific research and innovation technology in Sardinia''.

\end{document}